# Ultra-Compact Low-loss Broadband Waveguide Taper in Silicon-on-Insulator

**PURNIMA SETHI,[1] ANUBHAB HALDAR,[2] AND SHANKAR KUMAR SELVARAJA[1*]**

[1]*Centre for Nano Science and Engineering (CeNSE), Indian Institute of Science, Bangalore, India.*
[2]*Department of Physics, University of Massachusetts Amherst, USA*
*\*shankar.ks@cense.iisc.ernet.in*

**Abstract:** A novel design of large bandwidth, fabrication tolerant, CMOS-compatible compact tapers have been proposed and experimentally demonstrated in silicon-on-insulator. The proposed taper along with linear grating couplers for spot size conversion exhibits no degradation in the coupling efficiency compared to a focusing grating in 1550 nm band. A single taper design has a broadband operation over 600 nm that can be used in O, C, and L-band. The proposed compact taper is highly tolerant to fabrication variation; 80 nm change in the taper width and 500 nm in taper length changes the taper transmission by <0.4 dB. The footprint of the device i.e. taper along with the linear gratings is ~ 250 µm$^2$; this is 20X smaller than the adiabatic taper and 2X smaller than the standard focusing grating coupler

## 1. Introduction

The strong light confinement in high index-contrast waveguide platform allows compact devices and circuits, enabling dense optical integration. Over the years, Silicon-On-Insulator (SOI) wafer technology has emerged as a standard for realizing Complementary Metal-Oxide-Semiconductor (CMOS) technology compatible high-density photonic integrated circuits [1]. High-density integration brings new challenges in circuit design and routing. The devices with different waveguide width should be connected through low-loss interfaces. When considering high-density circuit, it is essential to reduce the footprint of these waveguide transitions. Since the taper length depends on the starting and ending waveguide width, the transition between a grating coupler (GC) and a single-mode photonic waveguide in a SOI platform is one of the largest [2-13]. Several designs of adiabatic tapers involving linear [7], exponential [8] and parabolic [9, 10] profiles have been proposed. However, there is a tradeoff between the taper length and coupling efficiency due to the adiabatic transition [11, 12].

A complex non-adiabatic taper 15.4 µm long using multimode interference (efficiency ~70%) [13] and lens-assisted focusing tapers with lengths varying from 10 µm to 20 µm with a loss of about 1dB for TE and 5dB for TM mode [14] has been demonstrated. Discontinuous tapers with 90% efficiency for the fundamental quasi-TM mode [15] have been theoretically shown. However, the proposed structures are either difficult to fabricate, or suffer from low efficiencies and larger footprints.

A grating footprint of 10 ×10 µm is typically chosen to mode-match the grating field with an optical fiber [16-22]. The grating is then coupled to a waveguide through a 150-500 µm long adiabatic taper [11, 23]. Thus, the footprint of the spot-size converters based on linear GCs is limited by the length of the adiabatic taper. To reduce the footprint of the coupler, a compact focusing grating was proposed and widely used as well [24]. The focusing grating allows an eight-fold length reduction in the footprint (~18.5µm×28 µm) without performance penalty compared to a linear GC with an adiabatic taper. However, focused gratings require accurate fiber alignment, bandwidth and reflection [25]. Thus, it would be extremely advantageous to use a linear GC with a short taper for a compact light-chip coupling scheme. The challenge is to design a compact taper that has low-insertion loss, low-reflection and is broadband as well as robust to fabrication imperfections.

In this paper, we propose and experimentally demonstrate an ultra-compact taper between a linear GC and single mode Si waveguide. The proposed taper is defined using a quadratic sinusoidal function and is merely 15 µm long with an insertion loss as low as 0.22 dB at 1550 nm and a bandwidth >150 nm. A detailed theoretical and experimental analysis of the proposed taper is presented. Furthermore, the performance of the proposed taper is experimentally compared with adiabatic taper and focused GCs.

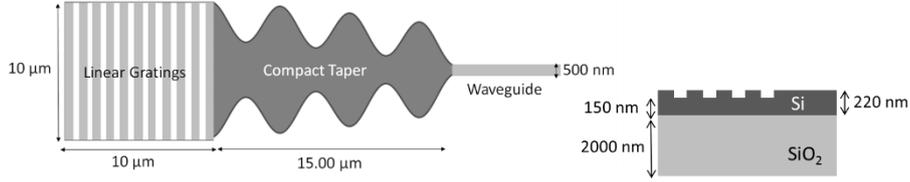

Figure 1. Schematic illustration of the proposed Compact Taper structure

## 2. Compact Taper: Design and Simulation

The schematic of the taper structure along with linear shallow-etched diffractive waveguide GCs is shown in Figure 1. Unlike an adiabatic taper, the proposed taper works on multi-mode interference along the length of the taper. The length and width of the taper are optimized to obtain interference progressively between the resonance modes along the taper resulting in maximum coupling to the fundamental waveguide mode. The proposed taper structure to connect a broad waveguide section to a submicron waveguide section is defined using an interpolation formula,

$$X = a\,(bz^2 + (1-b)z) + (1-a)\sin(c\,\tfrac{\pi}{2}\,z)^2$$
(1)

where $a$ is [0, 1], $b$ is [-1, 1], $c = \{2k + 1 : k \in Z\}$. This formula meets the following boundary conditions: $X\,(z = 0) = 0$ and $X\,(z = 1) = 1$ where z is the relative length of the taper. The final width profile, $X = f\,(z, a, b, c)$ is a superposition of a parabolic baseline and the square of a sine. The coefficient $a$ controls the fraction of the sinusoidal as well as the parabolic component. For case (i) $a = 0$, the sinusoids are at maximum amplitude, and the entirety of the taper is determined by the sinusoidal part of the function. For case (ii) $a = 1$, the sinusoidal component of the tapers is eliminated, and the taper follows a simple quadratic (or linear) change in width. The parameter $b$ controls the parabolic curvature of the baseline: case (i) $b = 1$, a convex parabola is obtained, case (ii) $b = -1$, a concave parabola is obtained, and case (iii) $b = 0$, a simple linear taper is obtained. Finally, $c$ controls the number of full oscillations of the sinusoidal component part of the taper. The restriction of $c$ (odd integers) is due to the boundary conditions that must be met at both ends of the taper's interpolation formula. All the three design parameters allow one to design an appropriate taper profile for maximum transmission between the waveguides.

Figure 2 illustrates the effect of the three design parameters $a$, $b$ and $c$ on the taper profile. A rigorous iterative optimization of these parameters was performed to identify suitable design parameters to obtain the shortest taper and high-transmission between the waveguide sections. In the following section, the proposed compact taper operating in the C and L band along with its performance metrics is presented. All the simulation was performed using finited difference and eigenmode expansion method.

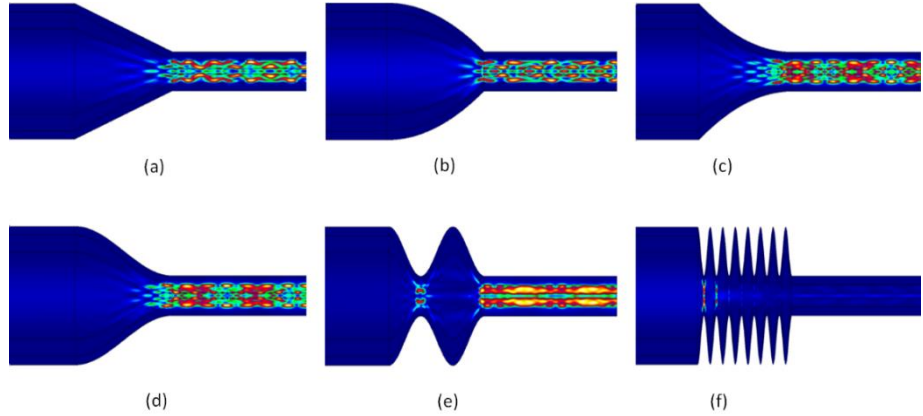

Figure 2. Optical Intensity Profiles of the Compact Taper at length = 14.50 μm for (a) $a = 1$, $b = 0$ (Linear Taper), Efficiency = 16%, (b) $a = 1$, $b = 1$, Efficiency = 25%, (c) $a = 1$, $b = -1$, Efficiency = 39%, (d) $a = 0$, $c = 1$, Efficiency = 38%, (e) $a = 0$, $c = 3$, Efficiency = 17%, (f) $a = 0$, $c = 15$, Efficiency = 9%.

An ultra-compact taper was designed in a 220/2000 nm Silicon/BOX SOI wafer technology. The taper was designed to couple a 10 μm (chosen to accommodate a linear GC) and a 500 nm wide waveguide. Ridge waveguide with an etch depth of 70 nm is used to keep the scattering losses low. As mentioned earlier, the taper design parameters $a$, $b$, $c$ and taper length was optimized to achieve maximum transmission at 1550 nm. Table 1 summarizes three taper configurations that yielded over 95% transmission efficiency. A maximum coupling efficiency of 96% is achieved for a taper length of 15 μm. We observed that $b$ values within the range of ~0.5-0.6 result in a coupling efficiency > 95% whereas a further increase in $b$-values i.e. 0.8-1.0 reduced the transmission to ~85%. Design values for $a$ and $c$ are fixed at 0.4 and 7 respectively. Figure 3 shows the corresponding optical intensity profiles for different optimized $b$-values and taper lengths.

Figure 4(a) depicts the spectral response of the compact taper-based linear GCs for the C + L band for all three configurations. The proposed taper has a broadband operation with the 3dB bandwidth of over 600 nm covering O, C, and L-band and beyond. Since the GCs operate in the C+L band, only results in this band are presented. Furthermore, the effect of dimensional variation on the transmission performance was also calculated to take fabrication tolerances into account. Figure 4(b) shows the effect of end waveguide width variation on the transmission. Since the taper was optimized for an end waveguide width of 500 nm, deviation results in a reduction in transmission. However, in practice one can expect a linewidth variation not more than 10% which corresponds to a width variation of ±25 nm. A variation in this range would result in transmission degradation by < 2% (0.08 dB), which shows the resilience of the proposed taper.

Figure 4(c) shows the effect of the total taper width variation on the coupling efficiency. The variation in taper width is obtained by varying the optimized $b$ value from 0.4 to 0.8. It is evident that the proposed structures have high manufacturing tolerances (> 88% efficiency when $b$ changes from 0.4 to 0.8 i.e. Δ shift in optimized taper width of 400 nm). For a change in b value from 0.50 to 0.58, there is a Δ change in Compact Taper's width of ~80 nm.

**Table 1.  Variation in Transmission for Different $b$ values at 1550 nm for Waveguide Width of 500 nm.**

| Configuration | a | b | c | Taper Length (µm) | Percentage Transmission | dB loss |
|---|---|---|---|---|---|---|
| CT1 | 0.4 | 0.50 | 7 | 14.50 | 95.03% | 0.22 |
| CT2 | 0.4 | 0.58 | 7 | 14.75 | 95.56% | 0.19 |
| CT3 | 0.4 | 0.62 | 7 | 15.00 | 96.14% | 0.17 |

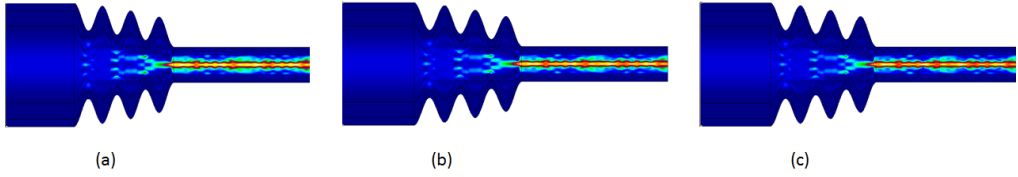

Figure 3. Optical Intensity Profile for the Compact Taper at 1550 nm TE polarization for different $b$ values and Taper Lengths (a) CT1, (b) CT 2, and (c) CT 3.

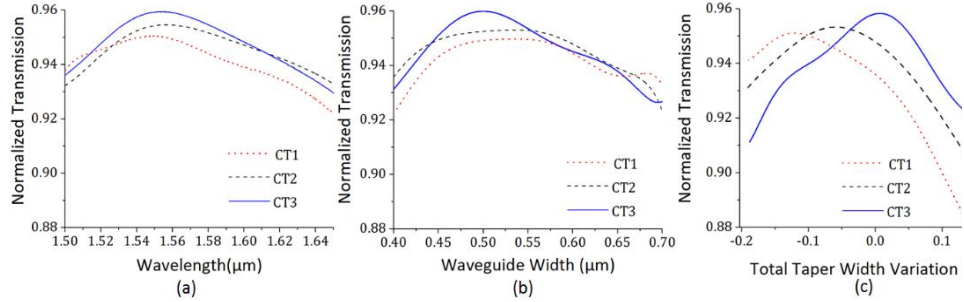

Figure 4. Spectral Response and Tolerance of the Proposed Compact Taper (Table 1), (a) Spectral Response of the Compact Taper in the C & L-band (1530 nm – 1625 nm), (b) Effect of End Waveguide Width Variation on the Transmission Efficiency of the Compact Taper (c) Effect of Compact Taper Width Variation (different $b$ values) on the Transmission Efficiency.

## 3. Experimental Results and Discussion

To compare the proposed taper performance with the existing designs, three type of configurations were fabricated; (i) linear GCs with adiabatic taper, (ii) focused GCs and (iii) the proposed Compact Taper with linear GCs. Figure 5 shows the schematic of the three tapers. The test structures were designed with an input GC with one of the tapers mentioned above coupling into a 500 nm ridge waveguide and taper-out to an identical output coupler configuration. Table 2 shows the combination of couplers-taper configurations and device specification. All the GCs were designed for TE-polarized 1550 nm with a grating period of 630 nm and 50% fill-factor. The design for the focused GC is adapted from the widely used Europractice PDK [26].

Table 2. Different Sets of the Devices Fabricated for C and L Band

| S. No. | Fiber to Waveguide Coupling | Waveguide to Fiber Coupling |
|---|---|---|
| 1 | Compact Taper (CT) | Compact Taper (CT) |
| 2 | Long Taper (500 µm) | Long Taper (500 µm) |
| 3 | Focused GCs | Focused GCs |

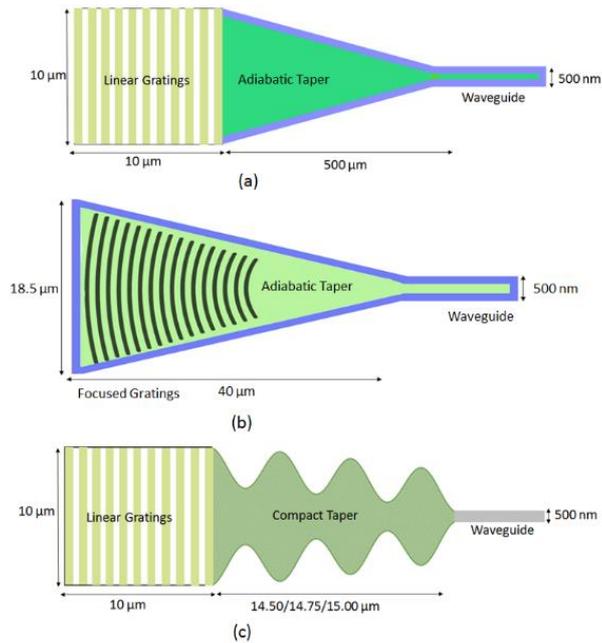

Figure 5. Schematic of the (a) Adiabatic Linear Taper along with Linear Gratings, (b) Focused Grating based Coupler using Curved Gratings, (c) Proposed Compact Taper.

The test structures were fabricated using electron-beam lithography and Inductively Coupled Plasma-Reactive Ion Etching (ICP-RIE) process. Pattering was done in a standard SOI substrate with a 220 nm thick device layer on a 2 μm buried oxide (BOX) layer. Figure 6 shows the SEM image of the proposed Compact taper.

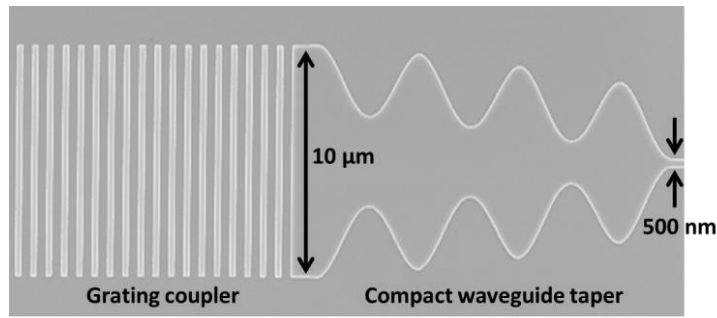

Figure 6. SEM image of a Compact Taper along with a linear grating coupler for 1550 nm TE polarization.

The fabricated devices were characterized using a tunable laser source (1510-1630 nm) and a photodetector. The polarization of the light from the laser source is controlled using polarization wheels before the input GC. The transmitted light is detected by an InGaAs photodetector. Figure 7 and Table 3 shows the summary of the characterization results. Figure 7(a) compares the performance of the Compact Taper with the Long Tapers and Focused GCs. Compact Taper configuration CT1 (Table-1) is used for the comparison. The characterization results show that the proposed CT1 yields the same coupling efficiency as a focusing GC and moreover provides a higher 3-dB bandwidth. The insertion loss per coupler is 5.45 dB, 6.1 dB and 5.3 dB for GC with Compact Taper, GC with adiabatic taper and focusing GC respectively. The insertion loss of the adiabatic long taper is about 1 dB higher, which we attribute to the waveguide loss in the adiabatic section. The 3-dB bandwidth which is another important performance metric for a GC is ~ 10 nm higher for CT1 compared to focusing GC.

Figure 7 (b) shows Compact Taper with different design parameter *b* and length. The variation in the design parameter *b* creates a taper waveguide width variation as illustrated in Figure 4(c). Measurement results show that a taper width variation of 80-160 nm and length variation of 500 nm would only vary the coupling efficiency by < 0.4 dB, which shows the robustness of the proposed taper. The proposed compact tapers are 20 times smaller (~ 250 μm$^2$) in comparison to linear GCs based couplers and 2 times smaller in comparison to focused GCs. Table 4 compares the performance of the three optimized configurations of the proposed compact taper.

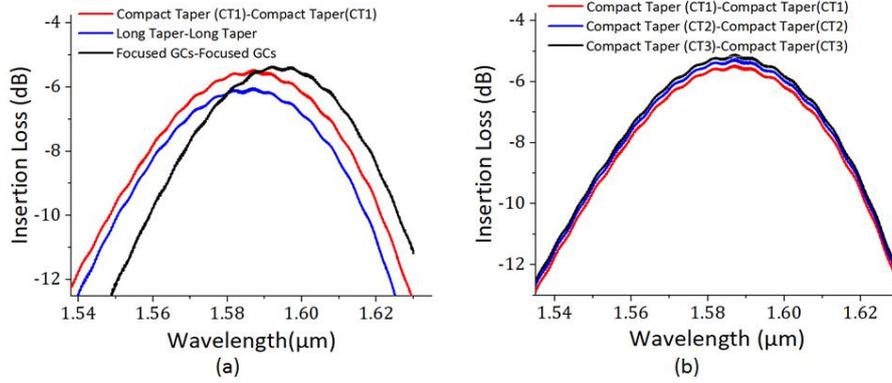

Figure 7. (a) Coupling Efficiency of the three configurations of Grating Coupler. (b) Effect of Design Parameters Variation on the Coupling Efficiency of the Proposed Compact Taper for three Variations as shown in Table 1.

Table 3. Experimental Analysis of the Performance-Metrics of the Various Configurations

| Set No. | Fiber to Waveguide Coupling | Waveguide to Fiber Coupling | Taper length [μm] | 3 dB Bandwidth (nm) | Loss per Coupler (dB) |
|---|---|---|---|---|---|
| 1 | Compact Taper | Compact Taper | 15 | 58.93 | 5.45 |
| 2 | Long Taper | Long Taper | 500 | 58.5 | 6.1 |
| 3 | Focused GCs | Focused GCs | 28 | 50.08 | 5.3 |

Table 4. Experimental Analysis of the Performance-Metrics of the Compact Taper Width Variation (different *b* values).

| Configuration | Loss per Coupler (dB) | 3 dB Bandwidth (nm) |
|---|---|---|
| CT1 | 5.45 | 58.93 |
| CT2 | 5.26 | 58.95 |
| CT3 | 5.12 | 58.99 |

## 4. Conclusion

Waveguide tapers are an essential part of a photonic integrated circuit, particularly, in Silicon the compact lateral waveguide tapers is necessary to realize coupling between devices of varying dimensions. We have designed and demonstrated the shortest tapered spot size converters to couple light to a single mode waveguide from a 10 μm wide waveguide. By using the taper with a linear GC, we have experimentally shown no degradation in coupling efficiency compared to standard focusing GC. We have also shown that by using a linear non-focusing grating, we achieve an improved 3dB bandwidth of ~59 nm against ~50 nm (focusing GC) in the 1550 nm band. The device shows a 20X reduction in the footprint of a single device based on linear GCs using adiabatic tapers and 2X reduction in comparison a focusing GC. We have also shown the fabrication tolerance of the compact taper. The proposed taper can be extended for use in other part for the circuits such as waveguide crossings.

**Acknowledgement**

The authors would like to thank the support from Defense Research and Development Organization, Government of India and Office of the Principal Scientific Advisor to Government of India. We also thank the staff of the National Nano-Fabrication Centre (NNFC) and the Micro and Nano Characterization Facility (MNCF) at the Indian Institute of Science-Bangalore for their assistance.